\documentstyle[12pt,aps,prd]{revtex}
\tightenlines
\input epsf

\newcommand{\beq}{\begin{equation}}
\newcommand{\eeq}{\end{equation}}

\def\lap{\lower.5ex\hbox{$\; \buildrel < \over \sim \;$}}
\def\gap{\lower.5ex\hbox{$\; \buildrel > \over \sim \;$}}
\def\lesssim{\lap}
\def\gtrsim{\gap}

\def\L{\Lambda}
\def\rL{\rho_\Lambda}

\begin{document}

\title{Solutions to the cosmological constant problems}

\author{J. Garriga$^{1,2}$ and A. Vilenkin $^2$}

\address{
$^1$ IFAE, Departament de F{\'\i}sica, Universitat Autonoma de Barcelona,\\
08193 Bellaterra (Barcelona), Spain\\
$^2$ Institute of Cosmology, Department of Physics and Astronomy,\\
Tufts University, Medford, MA 02155, USA}

\maketitle

\begin{abstract}

We critically review several recent approaches to solving the two
cosmological constant problems.  The "old" problem is the
discrepancy between the observed value of $\rL$ and the large
values suggested by particle physics models.  The second problem
is the "time coincidence" between the epoch of galaxy formation
$t_G$ and the epoch of $\L$-domination $t_\L$. It is conceivable
that the "old" problem can be resolved by fundamental physics
alone, but we argue that in order to explain the "time
coincidence" we must account for anthropic selection effects. Our
main focus here is on the discrete-$\L$ models in which $\L$ can
change through nucleation of branes.  We consider the cosmology
of this type of models in the context of inflation and discuss
the  observational constraints on the model parameters.  The
issue of multiple brane nucleation raised by Feng {\it et. al.}
is discussed in some detail. We also review continuous-$\L$
models in which the role of the cosmological constant is played
by a slowly varying potential of a scalar field. We find that
both continuous and discrete models can in principle solve both
cosmological constant problems, although the required values of
the parameters do not appear very natural. M-theory-motivated
brane models, in which the brane tension is determined by the
brane coupling to the four-form field, do not seem to be viable,
except perhaps in a very tight corner of the parameter space.
Finally, we point out that the time coincidence can also be
explained in models where $\Lambda$ is fixed, but the primordial
density contrast $Q=\delta\rho/\rho$ is treated as a random
variable.

\end{abstract}

\section{Introduction}

The cosmological constant $\Lambda$ presents us with at least two
intriguing problems. Particle physics models suggest that the
natural value for this constant is set by the Planck scale,
$M_{P}\sim 10^{18}$ GeV [we use the reduced Planck mass
$M_{P}=(8\pi G)^{-1/2}$]. The corresponding vacuum energy density
is $\rho_\Lambda\sim M_{P}^4$, which is some 120 orders of
magnitude greater than the observational bounds.  In
supersymmetric theories, one can expect a lower value, \beq
\rho_\Lambda\sim\eta_{SUSY}^4, \label{susy} \eeq where
$\eta_{SUSY}$ is the supersymmetry breaking scale. However, with
$\eta_{SUSY}\gtrsim 1$ TeV, this is still 60 orders of magnitude
too high. Until recently, this discrepancy between the expected
and observed values was the only cosmological constant problem.
Its solution, many believed, was that something so small could
only be zero, due to some unknown symmetry or dynamical
cancellation.

Thus, it came as a surprise  when recent observations
\cite{Supernova} provided evidence that the universe is
accelerating, rather than decelerating, suggesting a non-zero
cosmological constant\footnote{The surprise, however, was not
total. In Ref. \cite{AV95} (well before the supernova data
\cite{Supernova} would give the first observational evidence in
this direction) it was noted that anthropic selection effects
would place the cosmological constant in the range
$\rho_{\Lambda}/\rho_{M0}\lesssim 10$, and that "the actual value
is likely to be comparable to this upper bound." For a flat
universe this implies $\Omega_\Lambda \sim 0.9$, not far from the
observed value and certainly compatible with it, within the
accuracy of the prediction. Similar predictions where made in
\cite{Efstathiou} at about the same time.}. The observationally
suggested values of $\L$ correspond to $\rho_\L\sim\rho_{M0}$,
where $\rho_{M0}$ is the present density of matter.  This brings
yet another puzzle. It is difficult to understand why we happen
to live at the epoch when $\rho_M\sim\rho_\L$.  Another statement
of the problem is why the time when $\Lambda$ starts dominating
the universe nearly coincides with the epoch of galaxy formation,
\beq t_\L\sim t_G. \label{tLtG} \eeq This is the so-called cosmic
coincidence problem.

A number of proposed solutions to these problems have recently
appeared in the literature \cite{GV,Bousso,FMSW,Arkani,k-essence,Donoghue}.
Some of them rely on some form of the
anthropic principle, while others do not.  To our knowlege,
the only approach that can explain both puzzles is the one that attributes
them to anthropic selection effects.  In this approach, the
cosmological constant is assumed to be a random variable that can take
different values in defferent parts of the universe.

The purpose of this paper is to give a critical analysis of the
proposed approaches, both anthropic and otherwise.  Our main focus
will be on the models with a discrete spectrum of $\L$ which have
recently attracted much attention. We shall consider these models
in the framework of inflationary cosmology and discuss the
calculation of the probability distribution for $\rL$, as well as
the observational constraints on the model parameters.

The paper is organized as follows. In Section II we review the
motivation for considering $\Lambda$ as a random variable. In
Section III we discuss models where $\Lambda$ is a discrete
variable, in particular the models where there is a four-form
contribution to the cosmological constant, which may relax to a
small value through nucleation of branes. In Section IV we
analyze the cosmology of such models. In Section V we consider the
possibility of coincident brane nucleation. In Section VI we
discuss models where the cosmological constant is a continuous
variable. In Section VII we consider the possibility of a slowly
varying four-form field in theories with extra dimensions. In
Section VIII we review some non-anthropic approaches to the
problem. In Section IX we consider models where the time
coincidence is explained by assuming that the primordial density
contrast $Q=\delta \rho/\rho$ (and not necessarily $\Lambda$) is
a random variable. Our conclusions are summarized in Section X.

\section{$\Lambda$ as a random variable}
Not all values of $\Lambda$ are consistent with the existence of
conscious observers.  This observation was made by Barrow and
Tipler \cite{bt} (see also \cite{Davies}), but the first
quantitative analysis is due to Weinberg \cite{Weinberg87}. In a
spatially flat universe with a cosmological constant,
gravitational clustering effectively stops at $t\sim t_\L$.  At
later times, the vacuum energy dominates and the universe enters
a de Sitter stage of exponential expansion. An anthropic bound on
$\rho_\L$ can be obtained by requiring that it does not dominate
before the redshift $z_{max}$ when the earliest galaxies are
formed. Weinberg took $z_{max}\sim 4$ and obtained \beq
\rho_\L\lesssim 100\rho_{M0}. \label{Wbound} \eeq This is a
dramatic improvement over Eq.(\ref{susy}), but it still falls
short of the observational bound by a factor of about 30.

The anthropic bound (\ref{Wbound}) specifies the value of $\rho_\L$
which makes galaxy formation barely possible.  However, as it was
pointed out in \cite{AV95,Efstathiou},
the observers are where the galaxies are, and
thus most of the observers will detect not these marginal values, but
rather the values that maximize the number of galaxies.  More
precisely, the probability distribution for $\rho_\L$ can be written
as
\beq
d{\cal P}(\rL)={\cal P}_*(\rL)\nu(\rL)d\rL.
\label{dP}
\eeq
Here, ${\cal P}_*(\rL)d\rL$ is the {\it a priori} distribution, which
is proportional to the volume of those parts of the universe where
$\rho_\L$ takes values in the interval $d\rL$, and $\nu(\rL)$ is the
average number of galaxies that form per unit volume with a given
value of $\rL$.  According to the ``principle of mediocrity'', which
assumes that we are typical observers, Eq. (\ref{dP}) gives the
probability distribution for us to observe a given value of $\rL$.
The calculation of $\nu(\rL)$ is a standard
astrophysical problem; it can be done, for example, using the
Press-Schechter formalism \cite{PS}.  The {\it a priori} distribution
${\cal P}_*(\rL)$ should be determined from the theory of initial
conditions, e.g., from an inflationary model.

Martel, Shapiro and Weinberg \cite{MSW} (see also
\cite{Weinberg96}) presented a detailed calculation of $d{\cal
P}(\rL)$ assuming a flat {\it a priori} distribution, \beq {\cal
P}_*(\rL)=const \label{WC} \eeq in the range of interest
(\ref{Wbound}).  They found that the peak of the resulting
probability distribution is close to the observationally
suggested values of $\rL$. The cosmic time coincidence is easy to
understand in this approach \cite{GLV,Bludman}
if one notes that regions of the
universe where $t_\Lambda\ll t_G$ do not form any galaxies at all,
whereas regious where $t_\Lambda\gg t_G$ are suppressed
by "phase space", since they correspond to a very tiny range of
$\Lambda$. It was shown in Ref. \cite{GLV} that the probability
distribution for $t_G/t_\L$ is peaked at $t_G/t_\L \approx 1.5$,
and thus most observers will find themselves in galaxies formed
at $t_G\sim t_\L$.

This anthropic solution to the cosmological constant problems is
incomplete without a particle physics model that would allow $\L$
to take different values and without a theory of initial
conditions, such as an inflationary cosmological model, that would
allow one to calculate the {\it a priori} distribution ${\cal
P}_*(\rL)$.
%(The distribution ${\cal P}_*(\rL)$ does not have to be strictly
%constant, but a strong variation of ${\cal P}_*$ in the anthropic
%range would invalidate the argument.

One possibility is to consider models in which the role of the
vacuum energy is played by a slowly varying potential $V(\phi)$
of some scalar field $\phi$, which is very weakly coupled to
ordinary matter.  The values of $\phi$ are randomized by quantum
fluctuations during inflation, and analysis shows that the
resulting {\it a priori} distribution is indeed flat for a wide
class of potentials \cite{GV,Weinbergcomment}.  The main challenge
one has to face in this approach is to justify the
exceedingly flat potential $V(\phi)$ required by the model.
We shall comment on this issue in Section VI.
Before that, we shall consider an alternative possibility which
has recently attracted much attention. This is provided by models
with a discrete spectrum of $\rL$.

\section{Models with a discrete spectrum of $\L$}

The first model of this type was suggested in an early paper by Abbott
\cite{Abbott} as
an attempt to solve the old cosmological constant problem.  He
considered a self-interacting scalar field $\phi$ with a ``washboard''
potential $V(\phi)$ of the form illustrated in Fig.1.  The potential
has local minima at $\phi_n=n\eta$ with $n=0,\pm 1,\pm 2, ...$ ,
separated from one another by barriers.  The vacuum at $\phi=\phi_n$
has energy density
\beq
\rho_{\L n}= n\epsilon +{\rm const}
\label{washboard}
\eeq
and can decay through bubble nucleation to the vacuum at $\phi_{n-1}$.
%The bubble wall has thickness
%\beq
%\delta\sim V_0^{-1/2}\eta
%\label{deltaa}
%\eeq
%and tension
%\beq
%\sigma\sim V_0\delta\sim V_0^{1/2}\eta,
%\label{sigmaa}
%\eeq
%where $V_0$ is the height of the barrier.

\begin{figure}[t]
\centering \hspace*{-4mm}
\leavevmode\epsfysize=10 cm \epsfbox{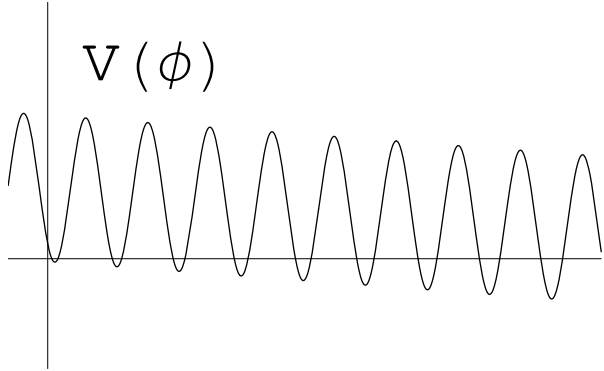}\\[3mm]
\caption[fig1]{\label{fig1}The washboard potential.}
\end{figure}

The nucleation rate $\Gamma_{n\downarrow}$ per unit spacetime
volume is given by \beq \Gamma_{n\downarrow}= A_n e^{-B_n},
\label{GammaAB} \eeq where $B_n$ is the action of the Coleman -
de Luccia instanton \cite{deLuccia} and the meaning of the
subscript $\downarrow$ will become clear shortly.  The bubble
radius at nucleation $R_n$ is bounded by $0<R_n<H_n^{-1}$, where
\begin{equation}
H_n^2={\rho_n \over 3 M_{P}^2} \label{Hn}
\end{equation}
is the square of the expansion rate of de Sitter space filled with
the vacuum $\phi_n$. The horizon radius and the curvature radius
of that space are both equal to $H_n^{-1}$.

An analytic expression for $B_n$ can be given in the thin wall
approximation, when $\delta\ll R_n$ \cite{deLuccia}. The general
expression is somewhat cumbersome and we shall only consider the
limiting cases of interest.  For $R_n\ll H_n^{-1}$, $B_n$ is
given by the flat space expression \cite{Coleman} \beq
B^{(flat)}_n\approx{27\pi^2\over{2}}{\sigma^4\over{\epsilon^3}},
\label{Bflat} \eeq approximately independent of $n$. In this
regime $R_n\approx 3\sigma/\epsilon$, so we should have $\sigma
H_n/\epsilon\ll 1$. In the opposite limit, $\sigma
H_n/\epsilon\gg 1$, we have $(H_n^{-1}-R_n)\ll H_n^{-1}$ and \beq
B^{(wall)}_n\approx 2\pi^2\sigma H_n^{-3}. \label{Bwall} \eeq The
vacuum energy difference $\epsilon$ is unimportant in this case,
and the action coincides with that for domain wall nucleation
\cite{Basu}.  The prefactor in Eq.(\ref{GammaAB}) can be
estimated as (see e.g. \cite{Jaume})
 \beq A_n\sim \sigma^2 R_n^2.
\label{A} \eeq
Eqs.(\ref{Bflat})-(\ref{A}) apply under the
condition that the gravitational effect of the wall is
negligible, \beq \sigma\ll M_P^2 H_n. \label{branegravity} \eeq

Upward quantum jumps from $\phi_{n-1}$ to $\phi_{n}$ are also
possible \cite{EWeinberg}. The corresponding nucleation rate is
\beq \Gamma_{(n-1)\uparrow}=\exp\left[{24 \pi^2
M_P^4}\left({1\over{\rho_n}} -{1\over{\rho_{n-1}}}\right)\right]
\Gamma_{n\downarrow}. \label{Gammaup} \eeq For $\epsilon\ll
\rho_n$ this can be approximated as \beq
\Gamma_{(n-1)\uparrow}=\exp\left(-{8\pi^2\over{3}}{\epsilon\over{H_n^4}}
\right) \Gamma_{n\downarrow}, \label{Gammaup1} \eeq where we have
used Eq.(\ref{Hn}) for $H_n$.

In order for the anthropic explanation to work, one needs
\beq
\epsilon\lesssim\rho_{M0}\sim (10^{-3}~{\rm eV})^4,
\label{epsilonbound}
\eeq
and in order to have successful baryogenesis, the energy density during
inflation should exceed $(1~TeV)^4$, which corresponds to
\beq
H\gtrsim 10^{-3}~eV.
\label{Hbound}
\eeq
Combining this with Eq.(\ref{Gammaup1}), we see that the probabilities
of upward and downward jumps in $\rL$ during inflation
are nearly equal, except perhaps in the borderline case when
\beq
H\sim\epsilon^{1/4}\sim 10^{-3}~{\rm eV}.
\label{borderline}
\eeq

An alternative discrete model, first discussed by Brown and
Teitelboim \cite{Teitelboim}, assumes that the cosmological
constant is due to a four-form field, \beq
F^{\alpha\beta\gamma\delta}={F\over{\sqrt{-g}}}\epsilon^{\alpha\beta\gamma
\delta}, \eeq which can change its value through the nucleation
of branes.  The total vacuum energy density is given by \beq
\rL=\rho_{bare}+F^2/2, \label{rhobare} \eeq where $\rho_{bare}<0$
is the `bare' cosmological constant at $F=0$.
%A vacuum with a large value of $\rL$ will decay through sequential
%nucleation and
%subsequent expansion of spherical branes.
The change of the field
strength across the brane is
\beq
\Delta F=\pm q,
\label{q}
\eeq
where $q={\rm const}$ is fixed by the model.
The four-form model has recently attracted much attention because
four-form fields with appropriate couplings to branes naturally arise
in the context of M-theory.  In this case the brane tension is
\cite{Bousso,FMSW}
\beq
\sigma=qM_P/\sqrt{2},
\label{sigma}
\eeq
and the effective thickness of the branes is $\delta\sim M_P^{-1}$, so
that the thin wall approximation is justified.

At present we should have $|F|\approx (-2\rho_{bare})^{1/2}$, so that
the bare cosmological constant is almost neutralized.
Then, in the range of interest, the spectrum of $\rL$ is nearly
equidistant, with a separation
\beq
\Delta\rL\equiv\epsilon\approx (-2\rho_{bare})^{1/2}q,
\label{epsilon}
\eeq
and the model is very similar to the Abbott's ``washboard'' model.
%In order for the anthropic explanation to work, one needs
%\beq
%\epsilon\lesssim \rho_{M0},
%\label{epsilonbound}
%\eeq
%which requires extremely small values of $q$ and $\sigma$.
We expect
\beq
|\rho_{bare}|\gtrsim (1 TeV)^4,
\label{rhobarebound}
\eeq
and it follows from
Eq.(\ref{epsilonbound}) that $q\lesssim 10^{-90}M_P^2$ and
\beq
\sigma\lesssim (10^{-3}~eV)^3,
\label{smallsigma}
\eeq
where we have used the relation (\ref{sigma}) between $\sigma$ and $q$.
Such small values of $q$ and $\sigma$ may appear problematic, but
in a recent paper \cite{FMSW} Feng, March-Russell, Sethi and Wilczek
(FMSW) have argued that they can naturally arise due to
non-perturbative effects in M-theory.  With $\sigma$ and $H$
satisfying the bounds (\ref{smallsigma}) and (\ref{Hbound}), the
condition of negligible brane gravity (\ref{branegravity}) is also
satisfied, and thus Eqs.(\ref{Bflat}),(\ref{Bwall}) can be used.

With the aid of Eqs.(\ref{sigma})-(\ref{smallsigma}) it can be easily
verified that the flat space bounce action (\ref{Bflat}) is bounded by
\cite{FMSW}
\beq
B^{(flat)}\lesssim 10^2.
\label{Bflatbound}
\eeq
This inequality is saturated for $\rho_{bare}\sim (1~{\rm TeV})^4$ and
\beq
\epsilon^{1/4}\sim\sigma^{1/3}\sim 10^{-3}~{\rm eV}.
\label{borderline2}
\eeq
If $\sigma$ and $\epsilon$ significantly differ from these borderline
values, then $B\lesssim 1$ and brane nucleation is unsuppressed.  A
similar bound is obtained for the wall nucleation action (\ref{Bwall})
using Eqs.(\ref{smallsigma}) and (\ref{Hbound}),
\beq
B^{(wall)}\lesssim 20.
\label{Bwallbound}
\eeq
Here, the inequality is saturated for
\beq
H\sim\sigma^{1/3}\sim 10^{-3}~{\rm eV}.
\label{borderline1}
\eeq
We note that Eqs.(\ref{sigma}),(\ref{epsilon}) apply
only to models based on M-theory, and therefore the constraints
(\ref{smallsigma}),(\ref{Bflatbound}), and (\ref{Bwallbound}) are also
limited to this class of models.

A different version of the four-form model has been developed by
Bousso and Polchinski (BP) \cite{Bousso}.  They assume that
several four-forms $F_i$ are present, so that Eq.(\ref{rhobare})
is replaced by \beq \rL=\rho_{bare}+{1\over{2}}\sum_i F_i^2. \eeq
The corresponding ``charges'' $q_i$ are not assumed to be very
small, but BP have shown that with multiple four-forms the
spectrum of the allowed values of $\rL$ can be sufficiently dense
to satisfy the condition (\ref{epsilonbound}) in the range of
interest. However, the situation here is quite different from
that in the FMSW model.  As pointed out by the authors
themselves, and further emphasized by Banks, Dine and Motl
\cite{Banks}, the vacua with nearby values of $\rL$ have very
different values of $F_i$ and are expected to have very different
physical properties.  There is no reason to expect the {\it a
priori} probabilities for these vacua to be similar. Moreover,
the low energy physics in different vacua is likely to be
different, so the process of galaxy formation and the types of
life that can evolve will also differ.  It appears therefore that
the anthropic approach to solving the cosmological constant
problems cannot be applied to this case \cite{Banks}.

\section{Cosmology of the four-form models}

\subsection{{\it A priori} distribution}

We shall now discuss the four-form models with brane nucleation in the
context of inflationary cosmology.  The energy density of the universe
during inflation can be expressed as
\beq
\rho_n(\chi)= U(\chi)+\rho_{\L n},
\eeq
where $U(\chi)$ is the potential of the inflaton field $\chi$,
$\rho_{\L n}$ is the cosmological constant contribution
(\ref{rhobare}), and index $n$ labels the vacuum energies
corresponding to different values of the
four-form field $F$.  [The inflaton potential is generally
$F$-dependent and has different forms $U_n(\chi)$ in defferent vacua
\cite{Bousso,Banks}.  Here we shall disregard this difference,
assuming that the variation of $U(\chi)$ is negligible in the narrow
anthropic range of $\rL$ that will be of interest to us.]
The minimum of $U(\chi)$ is assumed to be
at $U_{min}=0$.  The spacetime during inflation is locally
approximately de Sitter,
\beq
ds^2=dt^2 -e^{2H_n t}d{\bf x}^2,
\label{deSitter}
\eeq
with $H_n(\chi)$ given by Eq.(\ref{Hn}).
%We shall be mainly interested in the values of
%$\rho_{\L n}$ in the anthropic range (\ref{Wbound}), in which case $\rho_{\L
%n}\ll U(\chi)

A remarkable feature of inflation, which will play an important role
in our discussion here, is that generically
inflation never ends completely in the
entire universe.  The evolution of the inflaton field $\chi$ is
influenced by quantum fluctuations, and as a result thermalization
does not occur simultaneously in different parts of the universe.  In
most of the models, one finds that at any time there are parts of the
universe that are still inflating and that the total volume of
inflating regions is growing with time \cite{AV83,Linde86}.  This
picture is often referred to as stochastic, or eternal, inflation.

\begin{figure}[t]
\centering \hspace*{-4mm}
\leavevmode\epsfysize=10 cm \epsfbox{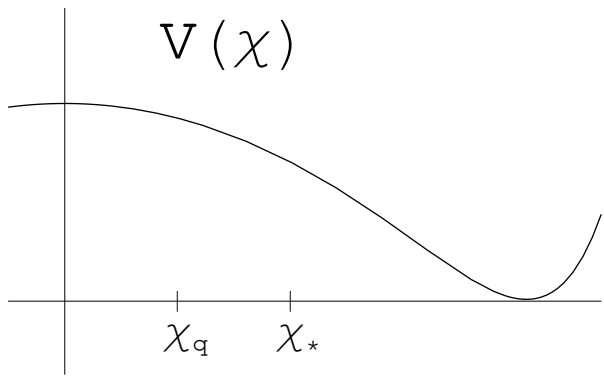}\\[3mm]
\caption[fig2]{\label{fig2} Inflaton potential}
\end{figure}

The full range of the field $\chi$ can be divided into the
``diffusion'', slow-roll, and thermalization parts, as
illustrated in Fig. 2. In the diffusion range,
$\chi\lesssim\chi_q$, the inflaton dynamics is dominated by
quantum fluctuations.  It is this regime that is responsible for
the eternal nature of inflation.  In the slow-roll regime,
$\chi_q\lesssim\chi\lesssim\chi_*$, the inflaton rolls down its
potential.  As it reaches the thermalization point $\chi_*$, it
starts oscillating about the minimum of the potential, and its
energy gets thermalized.  The hypersurfaces $\chi=\chi_*$ are
therefore the boundaries between inflating and thermalized
regions of spacetime. These surfaces play the role of the big
bang for the corresponding thermalized regions.  There is
typically an infinite number of such surfaces, each of them
having an infinite volume.  (For a discussion of the spacetime
structure of inflationary universe see, e.g., \cite{VVW}.)

As the inflaton $\chi$ fluctuates, rolls down its potential, and
eventually thermalizes its energy, spherical branes nucleate at the
rates (\ref{GammaAB}),(\ref{Gammaup}) changing the local values of the
four-form field $F$.  All possible values of $\rho_{\L n}$ will be
taken on each infinite thermalization surface $\Sigma_*$, and the {\it
a priori} probability ${\cal P}_{*n}$ can be defined as the fraction of the
volume of $\Sigma_*$ occupied by regions with vacuum energy density
$\rho_{\L n}$.  [The volume fraction on an infinite hypersurface can
be defined by calculating this fraction in a sphere of geodesic radius
$R$ and taking the limit $R\to\infty$.]

Brane nucleation can both decrease and increase the value of $\rho_{\L
n}$; the corresponding nucleation rates are related by
Eq.(\ref{Gammaup}).  For $\L$-lowering events, the bubble radius is
initially smaller than the horizon $H_n^{-1}$ and then grows in the
comoving coordinates, while for $\L$-raising events the radius is
initially larger than the horizon and then decreases in the comoving
coordinates.  In both cases, with an appropriate definition of the
nucleation time, the radius of the bubble nucleated at $t=0$
asymptotically approaches $H_n^{-1}e^{H_n t}$ \cite{GV98}.  This means
that the region affected by each nucleation event is a sphere of
initial radius $H_n^{-1}$.
For a comoving observer in vacuum $n$, the
probabilities per unit time to witness a $\L$-raising or lowering event
are
\beq
\kappa_{n \uparrow}=\Gamma_{n \uparrow}{4\pi\over{3}}H_n^{-3},
\eeq
\beq
\kappa_{n \downarrow}=\Gamma_{n \downarrow}{4\pi\over{3}}H_n^{-3}.
\eeq
It follows from (\ref{Gammaup}) that these probabilities are related
by
\beq
\kappa_{(n-1)\uparrow}=\kappa_{n\downarrow}(f_{n-1}/f_n),
\label{kapparel}
\eeq
where
\beq
f_n=H_n^{-3}\exp\left(-{24\pi^2M_P^4\over{\rho_n}}\right).
\label{fn}
\eeq

Consider an ensemble of comoving observers and let $p_n(t)$ be
the fraction of observers in the $n$-th vacuum, where $t$ is the
proper time along the observers' world lines.  The time evolution of
$p_n$ is described by the equations
\beq
dp_n/dt=-(\kappa_{n\uparrow}+\kappa_{n\downarrow})p_n
+\kappa_{(n-1)\uparrow} p_{n-1} +\kappa_{(n+1)\downarrow}p_{n+1}.
\label{pneq}
\eeq
Let us assume for a moment that the inflaton potential remains
unchanged,
\beq
U(\chi)={\rm const},
\label{Uconst}
\eeq
so that $\kappa_{n\uparrow}$ and
$\kappa_{n\downarrow}$ do not change with time.  Then the solutions of
(\ref{pneq}) approach the stationary distribution
\beq
p_n\propto f_n^{-1}\propto H_n^3\exp\left({24\pi^2M_P^4\over{\rho_n}}\right).
\label{pn}
\eeq
We shall be mostly interested in the probability distribution in the
anthropic range (\ref{Wbound}), where $\rho_{\L n}$ can be
approximated by (\ref{washboard}) with $\epsilon$ from
Eq.(\ref{epsilon}), and Eq.(\ref{pn}) takes the form
\beq
p_n\propto \exp\left( -{8\pi^2\epsilon\over{3H^4}}n\right).
\label{pn1}
\eeq
If inflation is well above the electroweak scale, $H\gg 10^{-3}~{\rm
eV}$, then the distribution (\ref{pn1}) is nearly flat in the anthropic
range,
\beq
p_n\approx {\rm const}.
\eeq

The assumption (\ref{Uconst}) may or may not be a good
approximation, depending on the shape of the potential $U(\chi)$.  A
simple example of a model where this approximation is adequate is a
``new inflation'' type model with a very flat potential in the
diffusion range near the maximum of $U(\chi)$ and a relatively steep
decline to the minimum in the slow roll range.  The distribution
(\ref{pn}) is established during the very long diffusion period, and
then it does not change much during the slow roll period if the
duration of the slow roll is shorter than the characteristic bubble
nucleation time.  Here we shall assume that the approximation
(\ref{Uconst}) is justified.

Can the distribution (\ref{pn}) be identified with the {\it a priori}
probability distribution ${\cal P}_{*n}$?  The answer is ``Yes, but
only in a restricted class of models''.  An ensemble of comoving
observers gives a comoving-volume distribution for $\rho_{\L n}$,
which does not account for the fact that regions with different values
of $\rho_{\L n}$ expand at different rates.  The condition for this
effect to be negligible is that brane nucleations should reshuffle the
values of $\rho_{\L n}$ between different regions on a timescale
$\tau_B$ which is much shorter than the time $\tau_H$ it takes for the
differential expansion rate to significantly modify the distribution,
\beq
\tau_B\ll\tau_H.
\label{tautau}
\eeq

As we noted in Section II, the probabilities of upward and
downward jumps in $\rho_\L$ should be nearly equal, except
perhaps in the borderline case (\ref{borderline}).  This means
that the evolution of $\rL$ can be pictured as a random walk with
steps taken on a timescale $\tau\sim\kappa^{-1}\sim
H^3\Gamma^{-1}$.  The anthropic range (\ref{Wbound}) comprises
$N\sim 10^2\rho_{M0}/\epsilon$ steps, and thus \beq \tau_B\sim
N^2 H^3 \Gamma^{-1}\sim 10^4\left({\rho_{M0}\over{\epsilon}}
\right)^2{H^3\over{\sigma^2 R_0^2}}e^B, \label{tauB} \eeq where
we have used Eqs.(\ref{GammaAB}) and (\ref{A}).  [In this
discussion we have dropped the subscripts $\uparrow$ and
$\downarrow$, since the upward and downward nucleation rates are
nearly equal, and the subscript $n$ since $H_n$ is nearly
constant in the anthropic range.]

The variation of the expansion rate in the anthropic range of $\rL$ is
\beq
\delta H\sim {N\epsilon\over{M_P^2 H}},
\eeq
and the time $\tau_H$ can be estimated as
\beq
\tau_H\sim 1/\delta H\sim 10^{-2} M_P^2 H/\rho_{M0}.
\label{tauH}
\eeq
The condition $\tau_B\ll\tau_H$ can now be expressed as
\beq
e^B\ll 10^{-6}{\epsilon^2\sigma^2 M_P^2 R_0^2\over{H^2 \rho_{M0}^3}}.
\label{tauBtauH}
\eeq
Parameter values satisfying this condition can be readily found.

%Using the relations $\epsilon\lesssim\rho_{M0}$, $R_0<H^{-1}$, and the
%bound (\ref{smallsigma}) on $\sigma$, we obtain
%\beq
%e^B\ll 10^{-6}{\sigma^2 M_P^2\over{H^4\rho_{M0}}}\lesssim 10^{32},
%\label{32}
%\eeq
%and thus
%\beq
%B\lesssim 10^2.
%\label{Bsmall}
%\eeq
%Note that the bound on $\sigma$ (\ref{smallsigma}) was derived from
%the relation (\ref{sigma}) which applies only to M-theory-inspired
%models.  Models unrelated to M-theory, such as the Abbott's washboard
%model, are therefore free from the constraint (\ref{Bsmall}).

What happens in the opposite limit, when $\tau_B\gg\tau_H$?  In this
case the differential expansion is important and the probabilities for
faster expanding regions with higher values of $\rL$ are strongly
enhanced.  The predicted values of $\rL$ should therefore be
significantly higher than those obtained with a flat {\it a priori}
distribution.  Martel {\it et.al.} \cite{MSW} have found that in the
latter case the probability distribution is peaked at
$\rL=0$, the width of the peak being somewhat broader than the
observationally suggested value.
Models with $\tau_B\gg\tau_H$ will have the peak displaced towards
higher values of $\rL$ and are therefore unlikely to give a good agreement
with observations.  A quantitative analysis of probability
distributions in such models can be given by a
relatively straightforward generalization of the formalism developed
in Ref. \cite{GV98}.

We note finally that in models with borderline values of
parameters (\ref{borderline}) the {\it a priori} distribution
(\ref{pn1}) can significantly deviate from flatness, with smaller
values of $\rL$ being favored. This would displace the peak of the
resulting distribution to negative values of $\rL$ and if
anything would make the observational situation even worse.

\subsection{Observational constraints}

Models of the type we are discussing here
suggest that we live in a
bubble surrounded by an expanding brane.  The values of $\rL$ inside
and outside the brane are different.
%If the bubble radius is smaller
%than the present horizon, $R\lesssim 3t_0$, then, assuming that we are not at
%the center of the bubble, we could see its effect in the
%anisotropy of the cosmic microwave background.  This effect would have
%been detected, unless the difference of vacuum energy densities across
%the brane is very tiny, $\epsilon\lesssim 10^{-5}\rho_{M0}$.
Let us first assume that the visible universe is contained within
a single bubble.  This means that the brane surrounding our
bubble nucleated before the presently observable universe crossed
the horizon during inflation.  For this situation to be typical,
the brane nucleation rates should be rather low, both during
inflation and at present.  This requires that the corresponding bounce
actions should be large, $B\gg 1$.  In
M-theory motivated models this is possible only for the
borderline values of the parameters, \beq
H\sim\epsilon^{1/4}\sim\sigma^{1/3}\sim 10^{-3}~{\rm eV}.
\label{borderline3}
\eeq
However, as we discussed at the end of Section IV.A, these values seem to be
disfavoured by observations.

The brane nucleation rate at present is given by
Eq.(\ref{GammaAB}) with $B$ and $A$ from
Eqs.(\ref{Bflat}),(\ref{A}).  In order to have no brane
nucleations in the observable universe in a Hubble time, we have
to require that \beq \Gamma t_0^4\lesssim 1, \label{Gammat0} \eeq
where $t_0$ is the present cosmic time.  For the parameter values
(\ref{borderline3}), $A\sim (10^{-3}~{\rm eV})^4$ and
Eq.(\ref{Gammat0}) gives $\exp(-B^{(flat)})\lesssim 10^{-116}$, or
\beq B^{(flat)}\gtrsim 270. \label{bigB} \eeq This is only
marginally consistent with the bound (\ref{Bflatbound})

The brane nucleation rate during inflation is determined by the
smaller of the two bounce actions (\ref{Bflat}),(\ref{Bwall}).
Eq.(\ref{borderline3}) tells us that in models based on M-theory brane
nucleation can be suppressed only if the expansion rate during
inflation is $H\sim 10^{-3}~{\rm eV}$.  Let ${\cal N}\sim 30$ be the
number of e-foldings from the time when the comoving region
corresponding to the presently observable universe crossed the horizon
to the end of inflation.  Then the size of this region at the end of
inflation is $H^{-1}e^{\cal N}$.  In order to have no brane
nucleations in this region during this whole period, we have to
require
\beq
\Gamma H^{-4}e^{3{\cal N}}\lesssim 1.
\eeq
For the parameter values (\ref{borderline3}) this gives $B\gtrsim 90$,
again marginally consistent with
(\ref{Bflatbound}),(\ref{Bwallbound}).

We thus see that M-theory based four-form models  could in principle
provide a solution to the cosmological constant problems, but only if
inflation is at a TeV scale and $\sigma$ and $\epsilon$ are in the
tight corner of the parameter space (\ref{borderline3}).  With such
values of the parameters, the condition (\ref{tauBtauH}) can be
(marginally) satisfied. However, from (\ref{Gammaup1}
we then find a significant
bias towards $\Lambda$-lowering nucleation events, which would shift
the a priori distribution (\ref{pn1}) towards lower $\Lambda$. This would
result in a prediction near the lower anthropic bound
 $\rho_{\Lambda} \sim -\rho_{M0}$. The
bias towards $\Lambda$-lowering events might be compensated to some
extent by the differential expansion rate, which adds relative volume to
regions with high $\Lambda$. However, both effects are exponential,
and unless there is a conspiracy in the parameters of the model, the
differential expansion is likely to be either insignificant
or dominant. In the latter case, the {\em a priori} distribution would be
biased towards large $\Lambda$, and it would be likely to predict a
cosmological constant much larger than observed. In summary, it seems
difficult to obtain a flat {\em a priori} distribution even in the
range (46). Of course, the possibility cannot be excluded with our order
of magnitude estimates, and there may still be a small viable region
of parameter space in this borderline range. We note also that for models
unrelated to M-theory the allowed parameter space is much larger.

Suppose now that the visible universe contains more than one
bubble. This would generally result in microwave background
anisotropies of amplitude $\delta T/T\gtrsim \epsilon/\rho_{M0}$,
so to avoid conflict with observations we have to require \beq
\epsilon\lesssim 10^{-5}\rho_{M0}. \label{epsilon5} \eeq This
takes us far from the borderline values (\ref{borderline3}), and
thus the multiple bubble scenario cannot be realized in M-theory
based models.  For non-M-theory models, a suitable set of
parameters can be easily found by choosing $\sigma$ and $H$
sufficiently large, while keeping $\epsilon$ under the bound
(\ref{epsilon5}).

The multiple bubble scenario is feasible only if branes have
negligible interaction with ordinary matter.  Otherwise we would see
fireworks along the bubble boundaries, where the branes hit the
stars and where they hit one another.
However, the gravitational impact of the branes cannot be
avoided.  An observer outside an expanding spherical bubble does not
experience any gravitational force until he is hit by the brane.
While the brane passes through the observer, the part of his body
inside the brane will experience an acceleration $a=GM/R^2$ relative
to the part of the body still outside the brane.  Here,
$M=(4\pi/3)\epsilon R^3$ and $R$ is the bubble radius at the moment of
impact.  With $R\sim t_0$ and $\epsilon$ satisfying (\ref{epsilon5}),
we have $a\sim G\epsilon t_0\sim (\epsilon/\rho_{M0})t_0^{-1} \lesssim
10^{-12}~{\rm cm/s^2}$.  The relative speed developed during the
passage time $\Delta t\sim 10^{-8}~{\rm s}$ is $\Delta v\lesssim
10^{-20}~{\rm cm/s}$, and the corresponding displacement is much
smaller than the inter-atomic distance.  For a brane passing through a
Sun-like star, $\Delta t\sim 10~{\rm s}$ and the displacement is still
smaller than the atomic scale.  Thus, if a brane is to sweep through
the Solar system, its only effect would be to set up imperceptible
vibrations in the objects it leaves behind.

What happens if $B<1$, so that brane nucleation is completely
unsuppressed?  The main danger here is that the vacuum energy will
decay so fast that it will drop significantly in less than a
Hubble time.  This can be countered by choosing $\epsilon$ so
small that the change in $\rL$ is negligible even after
nucleation of a large number of bubbles.  This case, however, is
almost indistinguishable from that of a scalar field with a very
flat potential, which will be discussed in Section VI.

\subsection{No empty universe problem}

Here we shall comment on the so called empty universe problem
which was encountered in all earlier work on discrete $\Lambda$
models \cite{Banks1,Abbott,Teitelboim,Bousso,FMSW,Donoghue,Banks}.  The
scenario these authors had in mind is that the universe starts
with a large cosmological constant and relaxes, within the
available cosmic time, to a metastable vacuum with an
observationally acceptable value of $\Lambda$.  The problem is
that, in order to make the present vacuum sufficiently stable,
the brane nucleation has to be strongly suppressed.  One then
finds that the time it takes the universe to evolve to the
low-energy vacuum is so large that, by the time when the process
is complete, any matter that the universe initially had gets
diluted to an extremely low density.  So one ends up with an empty
universe dominated by the cosmological constant.

A number of solutions to this problem have been proposed.
FMSW suggested \cite{FMSW} that the nucleation
rate of multiple coincident branes may be enhanced due to the
increased density of states.  They argued that this would lead to a
rapid descent of the vacuum energy towards lower values.  To ensure the
long lifetime of the present vacuum, they argued that this rate
enhancement may not apply to the vacuum with the lowest positive value
of $\rL$.  Bousso and Polchinski \cite{Bousso}, who considered brane
nucleation with large jumps in $\rL$, suggested that the penultimate
vacuum could have a high energy density.  The inflaton field would
then be excited to high values of its potential by quantum
fluctuations.  When the ultimate brane nucleates, the inflaton rolls
down the potential thermalizing its energy and providing a high
density of matter.  Alternatively, they suggested that the nucleation
of the ultimate bubble, which in their model is accompanied by a large
change in the four-form field $F$, can be accompanied by a large
modification of the inflaton potential.  As a result the inflaton will
be displaced from the minimum of the potential, even if it was at the
minimum prior to the bubble nucleation.

In our view, the empty universe problem is not a real problem, and
the attempts to solve it seem therefore unnecessary.  The problem
disappears when the eternal nature of inflation is taken into account.
As the inflaton fluctuates back and forth in the quantum diffusion
regime, branes are constantly being nucleated and all possible values
of $\rL$ are reached.  The slow rate of brane nucleation is not a
problem, since an unlimited amount of time is available.
Thermalization of the inflaton energy occurs at different times in
different parts of the universe, and each region inherits the local
value of $\rL$.  Each possible value is represented in the thermalized
regions of the universe.  We are interested only in those regions
where $\rL$ is in the anthropic range (\ref{Wbound}), because that's
where all the galaxies are.

\section{Multiple brane nucleation}

Up till now we assumed that brane nucleations change the four-form
field $F$ by a single unit, Eq.(\ref{q}).  However, nucleation of
multiple coincident branes is also possible. For $k$ coincident
branes there is a $U(k)$ super Yang-Mills (SYM) living on the
world-volume. In FMSW \cite{FMSW} it was argued that the
nucleation of coincident branes would be enhanced by a large
degeneracy factor
$$
D= e^{S},
$$
where $S$ is the "entropy" of the SYM fields. For 2-branes arising
from the wrapping of a $4D$-brane on a degenerating 2-cycle, FMSW
estimated this entropy as
\begin{equation}
S \sim k^{\beta} R^2 T^2. \label{entropy}
\end{equation}
Here $k^{\beta}$ counts the effective number of degrees of
freedom which live on the brane. There are theoretical
uncertainties in the exponent $\beta$, but FMSW suggest that it
should be between $2$ and $3/2$. $R$ is the radius of the
Coleman-De Luccia instanton, which coincides with the size of the
"bubble" at the time of nucleation, and $T$ is some effective
temperature. FMSW considered two different candidates for the
effective temperature. One of them was the effective ambient de
Sitter temperature \cite{gibbonshawking} $T_0$ before brane nucleation,
and the other was the geometric mean of $T_0$ and the effective temperature
$T_I$ of the new de Sitter space inside the nucleated brane $T
\sim (T_I T_0)^{1/2}$.

It is easy to understand, however, that the relevant effective temperature
corresponding to the Coleman-De Luccia (CdL) instanton is in fact
none of the above, but simply the effective de Sitter temperature
of the 2+1 dimensional world-volume of the brane
\begin{equation}
T = {1\over 2\pi R}. \label{gh}
\end{equation}
This is the temperature experienced by the degrees of freedom living in
the wall (and it is in fact higher than $T_0$ and $T_I$).
The prefactor $D$ is a determinant arising from Gaussian
integration of perturbations around the instanton solution,
including all light degrees of freedom. Such determinants where
discussed in some detail in \cite{Jaume}. A scalar field of mass
$m$ living on the 2+1 dimensional world-volume gives a
contribution
\begin{equation}
D_s=e^{\zeta'(0)/2} \label{degen}
\end{equation}
where $\zeta(z)$ is the Zeta function of the scalar fluctuation
operator on the 3-sphere. Its derivative at the origin is given
by \cite{Jaume}:
\begin{equation}
\zeta'(0)= 2 \zeta'_R(-2)
% +Q(y)
%$$
%where
%$$
%Q(y)=
-y^2 \ln(\sin \pi y)+{2\over \pi^2}\int_0^{\pi y} x \ln(\sin x)
dx, \label{zetadegen}
\end{equation}
where $y^2= 1- m^2 R^2$ and $\zeta_R$ is the usual Riemann Zeta function
(this expression is valid for light fields, with $mR < 1$).

For instance, the contribution of a conformally coupled scalar
field can be obtained by taking $m^2=(3/4) R^{-2}$, which gives
$$
\zeta'(0)= 2 \zeta'_R(-2) - {1\over 4} \ln 2 + {7\over
8\pi^2}\zeta_R(3) \approx -0.1276
$$
Hence, the effective degeneracy factor contributed by a conformal
scalar field is given by
\begin{equation}
D_{conf.}\approx e^{-.0638} \approx 0.91 < 1. \label{sup}
\end{equation}
The first thing to note is that this factor is not an enhancement,
but a suppression. Hence, the determinant cannot simply be
thought of as the exponential of an entropy.

In fact, the CdL instanton is not a thermal instanton, but a zero
temperatue instanton. Thermal instantons for brane nucleation are
static and have the topology $S^2 \times S^1$ (rather than $S^3$),
where the $S^2$ is the $2D$-brane at fixed time and the $S^1$ is
the periodic Euclidean time. Thermal instantons do in fact exist
also in de Sitter space, but they have not received too much
attention because their Euclidean action is always larger than
that of the maximally symmetric CdL. For thermal instantons (in
flat or in de Sitter space) the determinantal prefactor is given
by $D=e^{-\Delta F/T}$, where $\Delta F= \Delta E-TS$ is the free
energy contribution of light degrees of freedom on the brane.
Such prefactors have been considered in \cite{Jaume2}. The free
energy consists of the vacuum energy $\Delta E$ (or Casimir energy
on the two-sphere) minus the product of the temperature times the
entropy. While the entropy is always positive, the sign of the
Casimir contribution is notoriously dependent on the type of
field. In fact, for thermal instantons in de Sitter the
temperature is always smaller or equal to the inverse of the
size of the bubble, and hence the sign of the free energy
contribution can easily be dominated by the Casimir contribution,
which can have either sign. Although as mentioned above the CdL
instanton is not thermal, this consideration may help clarify why
the prefactor $D$ need not represent an enhancement. Depending on
the field content it may represent a suppression.

Another thing to note about (\ref{sup}) is that it is independent
of $R$. Roughly speaking, this is consistent with the fact that
the effective temperature is $T\propto R^{-1}$. In general,
however, the degeneracy factor will depend on $R$ and on the mass
of the particle. For light minimally coupled scalars,
equations (\ref{degen}) and
(\ref{zetadegen}) give
$$
D_s \approx {e^{{\zeta_R}'(-2)}\over \pi^{1/2} m R} \quad (mR\ll 1).
$$
There can be a strong enhancement in the nucleation rate if there
are very light massless scalar fields. In the limit $m \to 0$ the
factor goes to infinity. This is because a massless scalar has a
normalizable zero mode on the sphere, corresponding to the
symmetry  $\phi \to \phi + const$. In this case, the zero mode
must be treated as a collective coordinate. The nucleation rate is
proportional to the range $\delta\phi$ of the field $\phi$,
because the bubbles can be nucleated with any average value of
the scalar field with equal probability \cite{Jaume}
$$
\delta D_s(m^2=0) = \lim_{m^2\to 0} [m D_s(m)] (\pi R^3)^{1/2}
\delta \phi = e^{{\zeta_R}'(-2)} R^{1/2} \delta \phi.
$$
As we shall discuss in some examples below, some scalars are
likely to pick up masses of order of the intrinsic curvature of
the 2+1 sphere, and for these $D_s$ is also independent of the
radius.

Let us briefly consider the field content on the brane. For
$k$ coincident 4D-branes in ten dimensions, the effective theory is
$U(k)$ SYM. This consists of a $U(k)$ gauge field plus $5(k^2-1)$
scalar degrees of freedom in the adjoint representation of
$SU(k)$ plus $5$ scalar singlets plus the corresponding fermionic
degrees of freedom.

If the branes are flat (as in the case when there is
no external four-form field), then the theory is supersymmetric
and all scalar degrees of freedom are massless. For the case of a
single brane, the five scalars represent the goldstone modes of
the broken translational symmetry. That is, they correspond to
transverse displacements of the brane. For the case of two
branes, there are 10 such displacements. Five of them correspond
to simultaneous motion of both branes. These are the singlets
under $SU(2)$. The rest are in the adjoint representation, and if
they acquire an expectation value they give mass to two of the
four gauge bosons. For instance, when the two branes move apart,
one of the adjoint scalars acquires an expectation value and two
of the gauge bosons get a mass, breaking the symmetry $U(2) \to
U(1)\times U(1)$.

The case of interest to us is not a flat brane, but a
$4D$-brane wrapped on a degenerating two cycle. The world-wolume of
the resulting
$2D$-brane in 4 Euclidean dimensions is not flat either, but forms
a 3-sphere of radius $R$. In this situation, we do not expect the
theory to be supersymmetric. This is just as well, since in order
for the instanton to make any sense, some of the adjoint
scalars must pick up masses at one loop. Otherwise the instanton
would have too many zero modes and too many negative modes. To
simplify, let us consider the 3-sphere in 4 non-compact
dimensions. For the case of a single brane, the transverse
displacements correspond to a scalar field of mass $m^2=-3
R^{-2}$ \cite{Covariant}. This scalar field has a single negative
mode, which is the constant $l=0$ mode. A negative mode is
precisely what is needed for an instanton to contribute to the
imaginary part of the vacuum energy, and hence to contribute to
false vacuum decay \cite{Coleman}. Also, there are four
normalizable zero modes, which are the spherical harmonics with
$l=1$. These correspond to the four space-time translational
modes of the instanton, which have to be treated as collective
coordinates. This Goldstone field gives a determinantal prefactor
of the form \cite{Jaume}
$$
D = {\sigma^2 R^2 \over 4} e^{\zeta'_R(-2)} \Omega,
$$
where $\Omega=VT$ is the spacetime volume.  The prefactor $A$ in the
nucleation rate (\ref{A}) is obtained after dividing by $\Omega$.

If there are 2 coincident branes, then in principle there would
be two such fields $\phi_1$ and $\phi_2$ corresponding to the
independent transverse displacements of the brane. However, only
the combination $\phi_+ = (\phi_1 + \phi_2)$ will correspond to a
singlet under $SU(2)$. The orthogonal combination $\phi_-
=(\phi_1 - \phi_2)$ will be in the adjoint. As mentioned above, if
the instanton with two coincident branes is to make any sense,
this combination must aquire a positive mass at one loop so that
there is a single negative mode, not two, and four normalizable
zero modes in total. In other words, in order for the instanton to
make sense, the branes must attract each other. If they repelled
each other or if they did not "interact", then the two brane
configuration would in fact be an accidental superposition of two
independent bubbles in the "dilute gas" of instantons. The mass
of $\phi_-$ can be estimated as follows. The mass of the gauge
field $A_{\mu}$ is given by $m_A (\phi_-) \sim M_{P}^2 f(d)$,
where $M_{P}$ is the Planck mass, $d$ is the distance between
branes and $f(d)\approx d$ for $d \gg M_{P}^{-1}$. This is
because this vector corresponds to fundamental strings stretching
from one brane to the other. For smaller distances, $d \lesssim
M_{P}^{-1}$, we may expect a milder behaviour for the mass, which
we may heuristically parametrize as a power $f(d)\approx d
(M_{P}d)^n$, with $n>0$. The canonical field is related to the
distance through $\phi_- \sim d \sigma^{1/2}$. Hence,
$$
m_A^2 \sim M_{P}^4 f^2(\sigma^{-1/2}\phi_-).
$$
On a flat brane, the effective potential induced by a gauge field
of mass $m_A$ is proportional to $m_A^3$. However, it can be shown
that on a sphere there is also a term of order $m_A^2 R^{-1}$
which will in fact dominate at small $m_A$. When these terms are
added to the tree level potential $-3 R^{-2}\phi_-^2$, the scalar
aquires a very tiny expectation value $<\phi_-> \sim (\sigma/R
M_{P}^4)^{1/2n} \sigma^{1/2} M_{P}^{-1}$. In the broken phase,
$\phi_-$ has a positive mass squared of order $m^2_{\phi_-} \sim n
R^{-2}$. The gauge bosons will in turn aquire imperceptibly tiny
masses $m^2_A \ll R^{-2}$.

To summarize, some of the scalars may get very large masses from
the wrapping on a degenerating cycle. These will decouple, and
presumably will not contribute to the degeneracy factor. Others,
corresponding to the relative position of the 2-branes in 3+1
dimensions will get masses of order $R^{-1}$, and hence will
contribute degeneracy factors of order one, just like the
conformally coupled field discussed above. For the massless or
nearly massles gauge fields the contribution to the degeneracy
factor will be of order one (the vectors have no zero modes on
the sphere, so unlike the case of a scalar, a tiny mass will not
cause a large degeneracy factor). Similar considerations could be
applied to fermions. Thus we expect the total prefactor
to be of order
\begin{equation}
D_{total} \sim \Omega \sigma^2 R^2 e^{a k^{\beta}} \label{harl}
\end{equation}
where $a$ is a numerical factor and $k^{\beta}$ is, as  in
Eq. (\ref{entropy}), an estimate of the
effective number of degrees
of freedom. For a flat brane at weak coupling, $\beta=2$, but as
argued by FMSW it could be lower for the wrapped brane.
Unfortunately, without going into a very detailed analysis (which
is outside the scope of this paper) we are unable to determine
the sign of the constant $a$. However, as argued above, this value
seems to be rather insensitive to the value of $R$ or to the
value of the ambient de Sitter temperature.

For $a<0$ the nucleation of multiple branes is suppressed and we
are back to the situation described in Subsection IV.B. For $a>0$
a disaster may occur because transitions into deep Anti-de Sitter
space through multiple brane nucleation seem to be unsuppressed
due to a large degeneracy factor. FMSW suggested that the disaster
could be averted by an anthropic argument. If the step $\epsilon$
in the vacuum energy is of the order of 3 in units of
$\rho_{M0}$, allowing the values ...,-2,1,4,... then the
stringent anthropic bound for a negative cosmological constant
$\rho_{\Lambda}\gtrsim -\rho_{M0}$ would tell us that the vacuum
energy is in fact the lowest allowed value in the list, that is
$\rho_{\Lambda}\sim \rho_{M0}$ (note that this argument requires a
certain adjustment of the step $\epsilon$ in order to explain the
observed value). However, there is another problem which is how to
explain the stability of this vacuum once it has been reached. In
the $FMSW$ scenario, the stability was attributted to the fact
that in the vacuum with $\rho_{\Lambda}\sim \rho_{M0}$ the
effective temperature of the brane would be so low that the
degeneracy factor is switched off. However, as we have seen, the
degeneracy factor is quite independent on the ambient de Sitter
temperatures and hence it does not seem to switch off. The same
mechanism that would enhance coincident brane nucleation from a
high energy vacuum, would cause the disastrous decay of ``our''
vacuum.

Finally, we note that even if a mechanism could be devised to
switch off the degeneracy factor, so that the present vacuum is
stable, the time coincidence $t_G \sim t_{\Lambda}$ would be left
unexplained by this approach (just as in the non-anthropic models
discussed in Section VIII). Also, the unsuppressed nucleation of
coincident branes seems to preclude eternal inflation, and even
if one may intuitively argue that the lowest anthropic value is
the most probable, the actual probability distribution for
positive $\rho_{\Lambda}$ seems hard to estimate.

\section{Scalar field with a very flat potential}

In this class of models, what we perceive as a cosmological
constant is in fact a potential $V(\phi)$ of a scalar field
$\phi(x)$.  The potential does not have a succession of minima as
in Abbott's washboard model, but its slope is assumed to be so
small that the evolution of $\phi$ is slow on the cosmological
time scale. This is achieved if the slow roll conditions
\begin{equation}
 M_P^2 V''\ll V \lesssim \rho_{M0}, \label{sr1}
\end{equation}
\begin{equation}
 M_P V' \ll V \lesssim \rho_{M0}, \label{sr2}
\end{equation}
are satisfied up to the present time (here it is assumed that any
"true" cosmological constant is also included in the potential
$V$.) These conditions ensure that the field is overdamped by the
Hubble expansion, and that the kinetic energy is negligible
compared with the potential energy (so that the equation of state
is basically that of a cosmological constant term.)  The field $\phi$
is also assumed to have negligible couplings to all fields other than
gravity.

Let us now suppose, as in the previous sections, that there was a
period of inflation driven by a different scalar field $\chi$.
During inflation, massless scalar fields are randomized by
quantum fluctuations, which cause their root mean squared value
to increase with time as $\Delta\phi \sim H (Ht)^{1/2}$, where
$H$ is the inflationary expansion rate. If we consider a field of
mass $m$, this effect competes with the classical drift down to
the bottom of the potential, and after some time of order $t\sim H
m^{-2}$ a stationary distribution with root mean squared
$\Delta\phi \sim H^2 m^{-1}$ is established. This can be
interpreted in terms of the Gibbons-Hawking temperature $T \sim
H$ of de Sitter space as the condition $V(\phi) \sim m^2 \phi^2
\sim T^4$. In this example, all field values $|\phi| \ll H^2/m$
would be almost equally probable after the end of inflation. This
discussion, however, assumes that inflation proceeds at almost
the same rate for all field values in the range considered. That
is, the differential expansion rate $\delta H \sim V(\phi)(H
M_P^2)^{-1}$ is ignored.

The case of interest to us is slightly more general because the
potential need not be quadratic, and also because we are not
necessarily interested in field values near $\phi=0$. Rather, we
are interested in field values for which the energy density is in
the anthropically allowed range
\begin{equation}
-\rho_{M0} \lesssim V(\phi) \lesssim 100 \rho_{M0}
\label{rangero}.
\end{equation}
The differential expansion rate $\delta H \sim V(\phi)(H
M_P^2)^{-1}$ will be negligible if the time $t \sim
(\Delta\phi)_{anth}^2 H^{-3}$ that it takes for the field to
fluctuate accross the anthropic range of $\phi$ corresponding to
(\ref{rangero}) is smaller than $(\delta H)^{-1}$ for the same
range. This requires
\begin{equation}
(\Delta \phi)_{anth} \ll {H^2 \over 10 \rho_{M0}^{1/2}} M_P.
\label{flatr}
\end{equation}
If this condition is not satisfied, then the a priori probability
for the field values with a higher $V(\phi)$ would be
exponentially enhanced with respect to the field values at the
lower anthropic end. This would result in a prediction for the
effective cosmological constant which would be too high compared
with observations. Therefore, in what follows, we shall demand
that our potential satisfies Eqs. (\ref{sr1}), (\ref{sr2}) and
(\ref{flatr}).

\subsection{Solving both cosmological constant problems.}

Consider a potential of the form \cite{GV},
\begin{equation}
V(\phi) = \rho_{bare} + {1\over 2}\mu^2 \phi^2 \label{both},
\end{equation}
where $\rho_{bare}$ represents the "true" cosmological constant.
If $\rho_{bare}$ and $\mu^2$ have opposite signs, then the
effective vacuum energy will be very small when
\begin{equation}
|\phi| \approx |2\rho_{bare}|^{1/2}|\mu|^{-1} \label{largephi}.
\end{equation}
 The anthropic range is given by $(\Delta\phi)_{anth} \sim
100 \rho_{M0} |\mu^2\rho_{bare}|^{-1/2}$. Then, conditions
(\ref{sr1}), (\ref{sr2}) and (\ref{flatr}) imply
\begin{equation}
10^3 {\rho_{M0}^{1/2} \over H^2}{\rho_{M0} \over
|\rho_{bare}|^{1/2} M_P}\ll|\mu|\ll {\rho_{M0} \over
|\rho_{bare}|^{1/2} M_P}.
\end{equation}
From the cosmic microwave background temperature fluctuations we
know that $H\lesssim 10^{-5} M_{P}$. This leaves a wide range of
possibilities for the value of the mass parameter,
\begin{equation}
\mu \sim (10^{-167} - 10^{-120}) {M_P^3 \over |\rho_{bare}|^{1/2}},
\label{smallmass}
\end{equation}
spanning some 47 orders of
magnitude. Provided that $\mu$ is in this range, the {\em a
priori} probability distribution ${\cal P}_*(\phi)$ for $\phi$
will be flat. The probability distribution for the effective
cosmological constant $\rho_{\phi} = V(\phi)$ is given by
$$
{\cal P}_*(\rho_{\phi}) = {1\over V'} {\cal P_*}(\phi),
$$
and it will also be very flat, since $V'$ is almost constant in
the anthropic range. As mentioned in Section II, a flat {\em a
priori} distribution for the effective cosmological constant in
the anthropic range entails an automatic explanation for the two
cosmological constant puzzles \cite{GLV,Bludman}.

\subsection{A small mass from instantons?}

The challenge in the scenario presented above is to explain the
small mass parameter (\ref{smallmass}). In Ref. \cite{GV} we
suggested that this can be achieved through instanton effects.
For instance, $\phi$ could be a pseudo-Goldstone, the phase of a
scalar field $\Phi=\eta e^{i\phi/\eta}$ which spontaneously breaks
a global $U(1)$ symmetry. Since global charge can be swallowed by
wormholes, a small mass term for the field $\phi$ will be
generated through gravitational instantons \cite{kimyong,sual}.

Another possibility is that the phase may have an "axion"
coupling of the form
\begin{equation}
{\alpha_s \over \eta} \phi \tilde F F, \label{axion}
\end{equation}
where $F$ is the field strength of a "hidden" gauge sector with
gauge coupling constant $\alpha_s$. The coupling (\ref{axion})
will give a small mass to the pseudoscalar $\phi$ through
instanton effects.

We should mention, however, that there may be certain limitations
in implementing this idea in the present context. Consider an
instanton-induced potential of the form
$$
V(\phi) = \rho_{bare} + \Lambda^4 \cos(\phi/\eta).
$$
In order to solve the cosmological constant problem we need
$$
\Lambda^4 \gtrsim \rho_{bare}.
$$
Combining this with the slow roll conditions (\ref{sr1}) and
(\ref{sr2}) we find
$$
\eta \gg M_P {\rho_{bare} \over \rho_{M0}}.
$$
Thus, the expectation value $\eta$ must be truly huge compared
with the Planck scale.

In usual axion models, the effective vertex (\ref{axion}) can be
obtained in the following way. The scalar field $\Phi$ has Yukawa
interactions of the form $h \Phi \bar\Psi \Psi$ with an "exotic"
fermion $\Psi$ (here $h$ is the Yukawa coupling constant). The
fermion in turn interacts with the non abelian gauge fields, and
the coupling (\ref{axion}) is generated at one loop. The mass of
the fermions in the broken phase is given by $m_{\Psi} \sim h
\eta$. In our case, this mass is extremely large (unless $h$ is
extremely small), and so we can hardly trust the field theory
model for generating (\ref{axion}).

Perhaps more worrisome is the effect of wormholes. For small
symmetry breaking scale $\eta\lesssim M_P$ the scale $\Lambda^4$
in the instanton potential is of order $M_P^4\ e^{-S}$, where
$S\sim M_P/\eta$ is the wormhole action \cite{kimyong,sual}. The
radius of the wormhole is given by $R\sim (M_P \eta)^{-1/2}$.
This radius approaches the Planck scale as $\eta$ approaches
$M_P$, and the process becomes unsuppressed. The instanton
calculation becomes unreliable for higher values of $\eta$, but
it is not clear what would prevent nonperturbative gravitational
effects from completely destroying the global symmetry.

Therefore, as mentioned above, the generation of a small mass
through instantons may not have a straightforward implementation
in the present context. Clearly, this issue deserves further
investigation (see e.g. \cite{sual}).

\subsection{A very flat potential from field renormalization}

Consider a potential of the form \cite{Weinbergcomment}
\begin{equation}
V(\phi)  = \rho_{bare} + M^4 f(\lambda \phi), \label{c1}
\end{equation}
where $M$ is a reasonable mass parameter and $f$ is a function of
order one with no large or small parameters. If the parameter
$\lambda$ in the argument of $f$ is chosen to be very small, then
$V(\phi)$ will be very flat.  In particular, the mass term of the
field $\phi$ which has two powers of $\lambda$ will be very small.
Weinberg suggested \cite{Weinbergcomment}
that perhaps the smallness of $\lambda$ can be attributed to a large
running of the field renormalization $Z_{\mu}$ from some fiducial
short distance scale $\mu$ to the large scales in which the
cosmological constant is relevant, $\mu\to 0$.

More generally, as noted in Ref. \cite{Donoghue}, the effective
Lagrangian for a scalar field $\psi$ at large distances will
include nonminimal kinetic terms,
\begin{equation}
{\cal L} = F^2(\psi) (\partial_{\mu} \psi)^2 - V(\psi) + ...
\label{nm}
\end{equation}
Here, $F$ plays the role of a field renormalization, which in fact
may depend on $\psi$, and we have omitted terms with more
derivatives of $\psi$. If $F$ is very large, then the field
redefinition
$$d\phi = F d\psi$$
will result in a very flat effective potential for $\phi$.

Take for instance $F=e^{\psi/M_P}$ and $V=\rho_{bare}+ (m^2/2) \psi^2$,
where $m$ is a not too large mass parameter (see below). After the
change of variables we obtain
$$
{\cal L} = (\partial_{\mu} \phi)^2 -\rho_{bare}- {1\over 2}m^2
M_P^2 [\ln(\phi/M_P)]^2 +...
$$
The effective potential is now very flat at large $\phi$. The slow
roll conditions (\ref{sr1}) and (\ref{sr2}) are satisfied for
$$
\phi \gtrsim \phi_{min} = M_P {m^2\over H_0^2} \ln {m\over H_0}.
$$
The antropic range $m^2 \psi^2 \sim |\rho_{bare}|$ will satisfy
this condition provided that $m\ll |\rho_{bare}|^{1/2}M_P^{-1}$.
Finally, the condition (\ref{flatr}) is easily satisfied by
chosing a sufficiently high Hubble rate during inflation $H^2 \gg
10^3 \rho_{0M}^{1/2} (\phi/\phi_{min})$.

Thus, starting from a Lagrangian (\ref{nm}) with fairly simple
functions $F$ and $V$ we have been able to satisfy all necessary
conditions to solve both cosmological constant problems. Of
course, one may wonder why $F$ should have exponential behaviour
when $V$ is only polynomial, and it would be good to find a well
motivated physical setup where this Lagrangian emerges in a
natural way.

\section{A slowly varying four-form?}

In theories with extra
dimensions, the four form field strength is dynamical above the
compactification energy scale.
Donoghue suggested \cite{Donoghue} that in the early
universe the four-form might take a continuum of different values
in different parts of the universe, and that it might get frozen
to these values as the universe cooled down below the
compactification scale.
However, it is easy to show that the effective cosmological
constant can vary from place to place only if the size of the
internal space is also variable. As a result, the effect of the
four form is more properly described as a contribution to the
effective potential for the radius modulus of the extra space.

Let the higher dimensional manifold be the product of a
four-dimensional spacetime ${\cal M}$ and an internal space
${\cal S}$,
\begin{equation}
ds^2 = g_{\mu\nu}(x) dx^\mu dx^{\nu} + \sigma_{ij}(x,y) dy^i\
dy^j, \label{metric}
\end{equation}
where $\mu, \nu= 0,...,3$ are the four-dimensional indices and
$i,j=1,...,n$ are the internal space indices. The field strength
takes the form, ${\bf F} = f(x,y) {\bf \omega_4}+ ...$, where
${\bf \omega_4}= \sqrt{g}\ (\wedge_\mu dx^{\mu})$ is the
four-dimensional volume form ($g=-\det g_{\mu\nu}$) and the
ellipsis denote terms with at least one internal index (these do
not behave as a four-form upon dimensional reduction). The
equations of motion reduce to
\begin{equation}
d{\bf G}=0, \label{dg0}
\end{equation}
where ${\bf G}=^*{\bf F}=f(x,y){\bf \omega_n}+...$ is the dual of
the field strength and ${\bf \omega_n}= \sqrt{\sigma}\ (\wedge_i
dy^i)$ is the volume form on the internal space ($\sigma=\det
\sigma_{ij}$). Consider two different points $x_1$ and $x_2$ on
the 4D manifold ${\cal M}$, and a curve $\gamma_{12}$ joining
them. Applying Stokes theorem to the "cylinder"
$\gamma_{12}\times {\cal S}$ (where ${\cal S}$ is the internal
space), and using the equations of motion (\ref{dg0}), we
immediately find that
\begin{equation}
\int_{S;x_1} {\bf G} - \int_{S;x_2} {\bf G}=\int_{{\cal S}\times
\gamma_{12}} d{\bf G} = 0, \label{stokes}
\end{equation}
for arbitrary $x_1$ and $x_2$. Assume that the internal metric
factorizes as
$$
\sigma_{ij}= e^{2\psi(x)} \tilde\sigma_{ij}(y).
$$
Then, Eq. (\ref{stokes}) implies
$$
\bar f(x)= f_0 \ e^{-n\psi(x)},
$$
where $\bar f(x)$ is the average of $f(x,y)$ over the internal
space and $f_0$ is a constant. Kaluza-Klein modes average to zero
on the internal space and do not contribute to $\bar f$. However
such modes are massive in the reduced theory and do not behave as
an effective cosmological constant. It follows that the
contribution of the four form to the effective cosmological
constant is
\begin{equation}
{1\over 2} F^2 \equiv {1\over 2\cdot 4!}\int_{{\cal S}}
F_{\mu\nu\rho\sigma}F^{\mu\nu\rho\sigma} \sqrt{\sigma}\ d^n y=
 {1\over 2} F_0^2\ e^{-n\psi(x)}, \label{fv}
\end{equation}
where $F_0^2= f_0^2\ \int_{\cal S}\sqrt{\tilde\sigma}\ d^n y =
const$.

In the dimensionally reduced theory, $\psi(x)$ is a four
dimensional scalar field, and (\ref{fv}) is just a contribution
to its effective potential $V(\psi)$. At the classical level,
there are two other such contributions, due to a bare higher
dimensional cosmological constant $\Lambda^{(4+n)}_{bare}$ and
due to the curvature of the internal manifold. Following Refs.
\cite{freund}, it is easy to show that in terms of the Einstein
frame metric $\bar g_{\mu\nu} = e^{n\psi}g_{\mu\nu}$, the
effective action takes the form
$$
S= {M_P^2\over 2} \int \sqrt{\bar g}\ d^4 x\left[\bar R - {n
(2+n) \over 2}\ \bar g^{\mu\nu}\partial_\mu \psi \partial_\nu \psi
- V(\psi)\right],
$$
with
$$
V(\psi)= \Lambda_0\ e^{-n\psi} + {F_0^2\over 2}\
e^{-3n\psi}-{K\over 2}\ e^{-(2+n)\psi}.
$$
Here $M_P^2= M_*^{2+n} V_0$ and $\Lambda_0=
\Lambda_{bare}^{(4+n)} V_0$, where $M_*$ is the higher
dimensional Planck mass, $V_0=\int_{\cal S} \sqrt{\tilde\sigma}
d^n y = const.$, and $K$ is defined by $\tilde R_{ij}= (K/n)
\tilde \sigma_{ij}$. In general, the potential may have a minimum
but this need not be near $V=0$. For $n>1$ one can adjust the
parameters $\Lambda_0$ and $F_0$ so that the minimum is at the
right height to fit observations , but this would be the usual
fine-tuning (for $n=1$ the minimum must have negative effective
cosmological constant, so this tuning is not possible). We may
also consider the possibility that the field $\psi$ is away from
the minimum, but slowly rolling so that the effective potential
$V(\psi)$ plays the role of an effective cosmological constant,
as described in Section VI. The problem is that if the slow roll
conditions are met, then $\psi$ would have a negligible mass and
would mediate long range interactions of gravitational strength,
which are ruled out by observations.

\section{Non-anthropic approaches}

Here we comment on some attempts to solve
the cosmic coincidence problem without resorting to the anthropic
principle.
In a recent paper \cite{Arkani} Arkani-Hamed {\it et. al.}
suggested an explanation to the approximate coincidence of several
cosmological timescales: the time of matter-radiation equality
$t_{eq}$, the time of $\L$-domination $t_\L$, and the time of galaxy
formation $t_G$.  They assume that the Planck scale $M_P$ and the
electroweak scale $M_w$ are the only relevant scales and argue that
the temperature at matter-radiation equality and the vacuum energy
should then be given by
\beq
T_{eq}\sim M_w^2/M_P,
\label{A1}
\eeq
\beq
\rL\sim (M_w^2/M_P)^4.
\label{A2}
\eeq
It follows immediately from (\ref{A1}),(\ref{A2}) that $t_{eq}\sim
t_\L$.  This coincidence should of course be understood in a very
rough sense, since the actual values of $t_{eq}$ and $t_\L$ in our
universe differ by a few orders of magnitude.
Now, assuming the density fluctuation amplitude
\beq
Q\equiv\delta\rho/\rho\sim 10^{-5},
\label{deltarho}
\eeq
and using a more accurate value for $t_{eq}/t_\L$, the authors
show from (\ref{A2}) that the epoch of galaxy formation is at
\beq
t_G\sim t_\L.
\label{A3}
\eeq

In our view, a relation like (\ref{A2}) may account for the smallness
of $\L$ and may even explain its observed value.  However, the cosmic
coincidence (\ref{A3}) would remain unexplained.  The time of
$\L$-domination is determined by the value of $\L$, while the epoch of
galaxy formation is determined by the amplitude of density
fluctuations $Q$.  Even if we explain the value of $\L$, we still have
to explain why the value of $Q$ is such that $t_G\sim t_\L$.
Moreover, the accuracy of a few orders of magnitude is not sufficient
to explain the cosmic time coincidence: observations indicate that the
coincidence (\ref{A3}) is accurate within one order of magnitude.

Another non-anthropic approach to solving the cosmic coincidence problem
involves $k$-essence, a scalar field with a non-trivial kinetic term
\cite{k-essence}.  $k$-essence has a positive effective pressure
during the radiation era and starts acting as an effective
cosmological constant with the onset of matter domination.  With a
suitable choice of parameters it dominates the universe at $t_\L\sim
t_G$.  However, one could also choose parameters to obtain $t_\L\ll t_G$
or $t_\L\gg t_G$.  This model can explain why $\L$-domination occurs at
$t>t_{eq}$, but it cannot account for the coincidence (\ref{A3}).

\section{Models with variable Q}

Several authors have recently expressed the view that the
anthropic principle can perhaps be applied to the cosmological
constant problem - but to nothing else! \cite{dreams,Banks}
For instance, Weinberg has
remarked \cite{Weinbergcomment} that we cannot explain the masses
and charges of the elementary particles by assuming that they
depend on the expectation values of scalar fields with very
flat potentials. These light fields would
couple to the elementary particles, and would have been
observed in collisions and decays.

While this remark may be true, we can still apply the anthropic
principle to variables which determine the large scale properties
of the Universe, and which generically fall into the category of
``initial conditions''. Examples of these are the amplitude of
primordial fluctuations $Q$ \cite{TR,GTV,GLV}, the density parameter
$\Omega$ \cite{VW,GTV}, or even the baryon asymmetry.
In the inflationary context, these
parameters depend on the path that the inflaton field takes in
going from the diffusion regime to thermalization. The inflaton potential
represented in Fig. 2 is one-dimensional, and there is
a single path from the top of the potential to the local minimum.
However, more generally, the inflaton has several
components, and there may be a continuum of paths from the diffusion
region to a given minimum. Even if the low energy particle
physics Lagrangian is the same in all thermalized regions, and
even if there are no exotic light degrees of freedom after
thermalization, these regions may start with different initial
conditions which will be more or less favourable to galaxy
formation.

Consider for instance \cite{GTV} a two component scalar field
$\chi=\chi_1+i\chi_2=|\chi| e^{i\Theta}$, with potential $V(\chi) =
(g_1\chi_1^2 + g_2 \chi_2^2)/2$. This potential produces
inflation for $|\chi|\gtrsim M_P$. However, the amplitude of
density perturbations $Q$ depends on the direction $\Theta$ of
approach to the minimum, $Q\sim m(\Theta) N(|\chi|) M_P^{-1}$.
Here $m^2(\Theta)=g_1 \cos^2 \Theta+g_2 \sin^2\Theta$ and $N\sim
|\chi|^2 M_P^{-2}\sim 60$ is the number of e-foldings from the
time the present Hubble scale first crossed the horizon until the
end of inflation. The minimum at $\chi=0$ will be
reached from different directions in different thermalized
regions, and therefore these regions will have a different value
of $Q$ as an initial condition. This example illustrates that $Q$
can easily be made into a random variable. In general, its {\em a
priori} distribution ${\cal P}_*(Q)$ (i.e. its volume
distribution at the time of thermalization) will not necessarily
be flat in the anthropically allowed range. For any given model,
this distribution can be calculated using the numerical methods of
Ref. \cite{VVW}. To proceed, however, we shall heuristically
parametrize it as
\begin{equation}
d{\cal P}_*(Q) \sim Q^{-\alpha} d \ln Q, \label{qprior}
\end{equation}
where $\alpha$ is a constant (this may not necessarily be a good
estimate for the particlular model presented above, but we shall
use it anyway for the sake of argument.)

We may now take a point of view which is ``complementary'' to the
one used in the preceeding sections. Let us assume that the
cosmological constant is truly a constant of order $M_w^8
M_P^{-6}$ determined from the fundamental theory (as assumed e.g.
in \cite{Arkani}), and that $Q$ is a random variable with prior
distribution (\ref{qprior}). If $\alpha>0$, then low values of
$Q$ will be favoured {\em a priori}. However, if $Q$ is too low,
galaxies will not have time to form before the time $t_{\Lambda}$
when the cosmological constant starts dominating. With this, we
would basically explain why $Q\sim 10^{-5}$ as well as the time
coincidence $t_G\sim t_{\Lambda}$. These arguments can be made
more quantitative in the following way. The probability
distribution for a galaxy to form at time $t_G$ is
given by
\begin{equation}
d{\cal P}(t_G) \propto {\cal P}_*(Q) {d\nu(t_G,t_\Lambda,Q)\over
dt_G}  d \ln Q\ dt_G. \label{ptg1}
\end{equation}
Here, $\nu(t_G,t_\Lambda,Q)$ is the fraction of matter that
clusters up to the time $t_G$ in a universe where the density
contrast at the time of recombination is $Q$ and where the
cosmological constant is such that it will start dominating at
the time $t_\Lambda$. This fraction can be easily estimated by
using the Press-Schechter approximation. In Ref. \cite{GLV}  it
was shown [see Eq. (27) of that reference] that after integrating
over $Q$ the probability distribution for $t_G$ is given by
\begin{equation}
d{\cal P}(t_G) \propto {d F^{\alpha} \over dx} dx \label{ptg2}
\end{equation}
where
$$
F(x) = {5\over 6}\left({1+x\over x}\right)^{1/2} \int_0^x {dw
\over w^{1/6}(1+w)^{3/2}},
$$
and
$$
x= \sinh^2(t_G/t_{\Lambda}).
$$
(Following \cite{GLV}, we are using the convention that
$t_\Lambda$ is the time at which $\Omega_\Lambda = \Omega_M
\sinh^2 1$, where $\Omega_i$ are the fractional densities of
cosmological constant and non-relativistic matter respectively.)
The distribution (\ref{ptg2}) is plotted in Fig. 3 for different
values of $\alpha$, and we see that for moderate values of
$\alpha$ it presents a rather prominent peak at $t_G\sim
t_\Lambda$, as expected from the general arguments above.

\begin{figure}[t]
\centering \hspace*{-4mm}
\leavevmode\epsfysize=10 cm \epsfbox{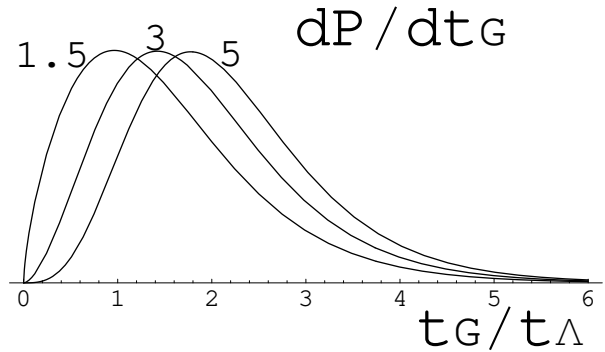}\\[3mm]
\caption[fig3]{\label{fig3} Probability distribution for $t_G/t_{\Lambda}$,
the time of galaxy formation as compared to the time when the cosmological
constant starts dominating. Here, $\Lambda$ is taken to be a
fundamental constant but the density contrast $Q$ is treated as a random
variable with {\em a priori} volume distribution $\propto Q^{-\alpha}$ at the
time of thermalization. The plot is shown for $\alpha=1.5,\ 3$ and $5$. The
distributions present rather well defined peaks at $t_G \sim t_\Lambda$.}
\end{figure}

Finally, one may take the view that both $\Lambda$ and $Q$ are
random variables. This possibility was considered in \cite{GLV},
where it was shown that a decreasing {\em a priori} distribution
for $Q$ pushes the cosmological constant to small values, so that
both $t_G$ and $t_\Lambda$ tend to be very large. In this case, a
new time scale comes into play. This is the so-called cooling
boundary $t_{cb}$ \cite{TR}. For times $t>t_{cb}$ gravitationally
collapsing clouds of galactic mass cannot fragment into stars because
they are too cold to reach the usual ``cooling'' line emission thresholds,
and they stay as pressure supported configurations for a very long time. Thus,
usual galaxy formation is suppressed after $t\sim t_{cb} \approx
3\cdot 10^{10} Yr$. This time is determined from microphysical
parameters such as the fine structure constant, the proton mass
and the fraction of baryonic matter. Since the time of galaxy
formation cannot be arbitrarily large, in the situation where
both $Q$ and $\Lambda$ are random variables we expect $t_G\sim
t_{\Lambda} \sim t_{cb}$ (see \cite{GLV} for details).
There are many uncertainties associated with the calculation of
$t_{cb}$. Perhaps after some of these uncertainties are removed, we may
actually find that $t_{cb}\gg t_G\sim t_{\Lambda}$.
This hypothetical situation would suggest that one of the timescales $t_G$ and
$t_\Lambda$ is not a random variable, or that if both of them
are, then their {\em a priori} distribution must have a rather
peculiar behaviour. This in turn would give us information on the
theories of initial conditions giving rise to these {\em a priori}
distributions.

These examples seem to suggest that the applicability of
anthropic reasoning, once it is accepted, may easily go beyond
the issue of the cosmological constant problem.

\section{Conclusions}

The anthropic principle has a bad reputation. It is often
regarded as a handwaving argument relying on poorly understood
phenomena like intelligent life and having no predictive power.
Although this criticism is not entirely ungrounded, there is a
class of cosmological models where the use of anthropic principle
is not only justified but may in fact be inevitable, and where it
can be used to make quantitative predictions. These are the
models in which some cosmological parameters, or physical
``constants'', take different values in different parts of the
universe.  In such models, one cannot predict the precise values
of the parameters that we are going to observe.  One can only
hope to calculate the corresponding probability distributions.
The criteria for justifying (and compelling) the use of anthropic
principle are that the model should provide (i) a mechanism for
variation of the parameters and (ii) a way of calculating the
probability distributions. Once the probabilities are calculated,
one can predict that the parameters are going to be observed
within a certain range of values, say, at a 95\% confidence
level.  This seems to be as quantitative as one can possibly get
in this class of models.

From a practical
point of view, parameters that we can hope to determine
anthropically should satisfy the condition that they do not
affect life processes, and preferably should also not affect the
poorly understood astrophysical processes such as star formation
\cite{VW,GLV,Banks}. So for example the gravitational constant may
be hard to determine anthropically with our present level of
understanding, since it affects both the evolution of life and
star formation. However, gravity does not change chemistry, which
is already a big simplification. So it is not inconceivable that
the value of Newton's constant may in the future receive an
anthropic explanation.

In this paper we discussed anthropic approaches to solving the two
cosmological constant problems (CCPs).  The first (old) CCP is the
discrepancy between the observed small value of $\rL$ and the
large values suggested by particle physics models.  The second
(time coincidence) CCP is the puzzling coincidence between the
epoch of galaxy formation $t_G$ and the epoch of $\L$-domination
$t_\L$.  While it is conceivable that the old CCP can be resolved
by fundamental physics \cite{Witten,Dvali,Arkani,Guendelman},
we have argued that the time
coincidence problem calls for an anthropic explanation.

We first considered models with a discrete spectrum of $\rL$.
These include Abbott's scalar field model with a "washboard"
potential \cite{Abbott}, as well as models in which $\rL$ can
change through brane nucleation accompanied by a change of the
four-form field $F$ \cite{Teitelboim,Bousso,FMSW,Donoghue}.
Such models can solve both
CCPs, provided that {\it (i)} the separation between the discrete
values of $\rL$ is $\epsilon\lesssim \rho_{M0}$, where
$\rho_{M0}$ is the present matter density, {\it (ii)} the
probability distribution for $\rL$ at the end of inflation is
nearly flat, ${\cal P}_*(\rL)\approx {\rm const}$, and {\it
(iii)} the brane nucleation rate is sufficiently low, so that the
present vacuum energy does not drop significantly in less than a
Hubble time.  We discussed the cosmology of this class of models,
the calculation of the prior distribution ${\cal P}_*(\rL)$, and
the observational constraints on the model parameters.

The required values of the "level separation" $\epsilon$ may
appear uncomfortably small, but Feng, March-Russell, Sethi and
Wilczek (FMSW) \cite{FMSW} have argued that they can naturally
arise due to non-perturbative effects in M-theory. In
M-theory-related models, the brane tension $\sigma$ is related to
$\epsilon$ through $\epsilon\sim \sigma\rho_{bare}^{1/2}/M_p$ and
should also be very small. Our analysis shows that in such models
the conditions {\it (i)-(iii)} cannot be satisfied without
fine-tuning of the parameters.

It was conjectured by Weinberg \cite{Weinberg87} that the
condition {\it (ii)} of a flat
{\em a priori} distribution for $\Lambda$
would automatically be satisfied in any particle physics
model where the cosmological constant is a random variable.
In Ref. \cite{GV} we showed that this conjecture is not always
satisfied in models where the role of the cosmological constant
is played by a slowly varying field. Here, we have shown that
the conjecture is generically not satisfied in four-form models
either. In fact, this condition has to be enforced in order to fit
observations. This, in turn, places
severe constraints on the model parameters. Hence, in trying to solve the
cosmological constant problems by anthropic means, the flat {\it a
priori} distribution for $\L$ cannot be taken for granted and the
problem of calculating ${\cal P_*}$ has to be addressed.

Bousso and Polchinsky \cite{Bousso} have studied models with
multiple four-form fields $F_i$ and found that the spectrum of
the allowed values of $\rL$ can be sufficiently dense even for
large brane tensions. However, in this case the vacua with nearby
values of $\rL$ have very different values of $F_i$, and a flat
probability distribution required in {\it (ii)} is rather
unlikely.  Moreover, the low-energy physics in such vacua is
likely to be different, and it appears that the anthropic
approach to solving the CCPs cannot be applied in this case
\cite{Banks}.

For models unrelated to M-theory, $\sigma$ and $\epsilon$ are
generally unrelated, and values consistent with the constraints
{\it (i)-(iii)} can easily be found.  However, if one gives up
the M-theory connection, then the FMSW argument cannot be used,
and one has to seek an alternative explanation for the tiny value
of $\epsilon$.  Alternatively, one might seek modifications of
the FMSW model that could relax the relation between $\epsilon$
and $\sigma$.

All of the earlier discussions of the cosmology of discrete
$\Lambda$ models encountered the "empty universe problem"
\cite{Banks1,Abbott,Teitelboim,Bousso,FMSW,Donoghue,Banks}.
In order to make the present vacuum sufficiently
stable, the brane nucleation has to be strongly suppressed.  One
then finds that the time it takes the universe to evolve from
some initial high value of $\rL$ to the present low value is much
greater than the present Hubble time.  This suggests that by the
time the process is complete, any matter that the universe
initially had may get diluted to an extremely low density, so
that one would end up with an empty universe dominated by the
cosmological constant.

We have argued that the empty universe problem disappears when the
eternal nature of inflation is taken into account.  During
inflation, brane nucleations leading to higher and lower values
of $\rL$ have nearly equal probabilities.  As a result, the
values of $\rL$ are randomized, with different parts of the
universe thermalizing with different values.  The resulting
probability distribution ${\cal P}_*(\rL)$ can be calculated
using the stochastic formalism we developed in Section IV.  The
slow rate of brane nucleation is not a problem in eternal
inflation, since an unlimited amount of time is available.

FMSW suggested an interesting possibility that nucleation of
multiple branes could be enhanced by a large degeneracy factor
due to the light fields living on the branes.  If true, this
could significantly modify the brane model cosmology.  In Section
V we studied multiple brane nucleation in some detail and found
that the pre-exponential factor in the brane nucleation rate can
both enhance and suppress multiple brane nucleation, depending on
the field content of the branes.  We also concluded that models
in which multiple brane nucleation dominates can be ruled out,
because in such models there is nothing to prevent our present
vacuum from tunneling down to deep anti-de Sitter space.

We also discussed models with a continuous spectrum of $\rL$, in
which the role of the cosmological constant is played by the
potential $V(\phi)$ of a scalar field $\phi(x)$.  The potential
has to be very flat, so that its value does not significantly
evolve on the present Hubble time scale.  The values of the field
$\phi$ are randomized by quantum fluctuations during inflation,
and models can easily be constructed in which the resulting
probability distribution for $V(\phi)$ is nearly flat in the
range of interest, thus solving both CCPs.  The challenge here is
to justify the very flat potentials required in this class of
models.  Possibilities include a pseudo-Goldstone field which
acquires a potential through instanton effects \cite{GV}, a
large running of the field renormalization \cite{Weinbergcomment}, and
a non-minimal kinetic term with an exponential $\phi$-dependence
\cite{Donoghue}.  We have pointed out some difficulties of the
instanton approach.

We thus see that both discrete and continuous $\Lambda$ models
could in principle solve both of the CCPs.  However, none of the
models that have been suggested so far appears particularly well
motivated or natural.

An alternative approach is to assume that the old CCP can be
solved within the fundamental theory.  The cosmological constant
is then truly a constant and is given by an expression like
$\rL\sim M_w^4 M_P^{-2}$, as in \cite{Arkani}.  At the same time,
the amplitude of density fluctuations $Q$ could be a random
variable, so that the epoch of galaxy formation $t_G$ is
different in different parts of the universe. We have shown in
Section VIII that, for a wide class of prior distributions ${\cal
P}_*(Q)$, most of the galaxies will be in regions where $t_G\sim
t_\L$, thus explaining the cosmic time coincidence. It would be
interesting to extend this analysis and calculate the
distribution ${\cal P}_*(Q)$ for some models with a variable
$Q$.  One would then have some idea of how naturally the
distributions of the required type can be obtained.

\section*{Acknowledgements}

J.G. is grateful to Alex Pomarol and to Klaus Kirsten for useful
and enjoyable discussions. This work was supported by the
Templeton Foundation under grant COS 253. J.G. is partially
supported by CICYT, under grant AEN99-0766, and by the Yamada
Foundation. A.V. is partially supported by the National Science
Foundation.

\section*{Notes added}

1- After this paper was submitted for publication, a revised version of
Ref. \cite{FMSW} has appeared. There, it is pointed out that
the relation (21) between the tension $\sigma$ and the charge density
$q$ of the brane does not hold for branes wrapping on
degenerating cycles.
Instead, the tension is suppressed by an exponential factor relative to
the charge.  We note two potential problems with this picture.  First, as
it was argued in Ref. \cite{giaalex}, the brane charge and tension appear to
be unprotected against renormalization below the supersymmetry breaking
scale.  Such renormalization would make the brane charge $q$ unacceptably
large.  Second, if for some reason the brane parameters do not get
renormalized, then, in order to satisfy the anthropic constraint (22) on
$q$, the brane tension has to be exceedingly small.  The instanton action
(9) would then be small and brane nucleation would be completely
unsuppressed.

2- A new approach to explaining very flat scalar potentials and branes
with a very small four-form charge has been suggested in Ref. \cite{giaalex},
where these features are attributed to a spontaneously broken discrete
symmetry.

\end{document}